\documentclass[nofootinbib,10pt,aip,twocolumn,showpacs]{revtex4-1}

\usepackage[english]{babel}

\usepackage{amsmath}
\usepackage{graphicx} 
\usepackage{multirow} 

\usepackage{hyperref}  
\usepackage{xcolor}
\usepackage{soul}

\graphicspath{{./figures/}} 

\begin{document}

\title{First-principles study of MoS$_{2}$ and MoSe$_{2}$ nanoclusters in the framework 
       of evolutionary algorithm and density functional theory} 
\author{Zohre Hashemi}
\thanks{These authors contributed equally to this work.}
\author{Shohreh Rafiezadeh}
\thanks{These authors contributed equally to this work.}
\author{Roohollah Hafizi}
\email{r.hafizi@ph.iut.ac.ir}
\author{S. Javad Hashemifar}
\author{Hadi Akbarzadeh}
\affiliation{Department of Physics, Isfahan University of Technology, 
             Isfahan, 84156-83111, Iran}

\begin{abstract}

Evolutionary algorithm is combined with full-potential ab-initio calculations
to investigate conformational space of (MoS$_2$)$_n$ and (MoSe$_2$)$_n$ ($n=1-10$) 
nanoclusters and to identify the lowest energy structural isomers of these systems.
It is argued that within both BLYP and PBE functionals,
these nanoclusters favor sandwiched planar configurations,
similar to their ideal planar sheets.
The second order difference in total energy ($\Delta_2$E) of the lowest energy isomers are computed 
to estimate the abundance of the clusters at different sizes and to 
determine the magic sizes of (MoS$_{2}$)$_n$ and (MoSe$_{2}$)$_n$ nanoclusters.
In order to investigate the electronic properties of nanoclusters, 
their energy gap is calculated by several methods,
including hybrid functionals (B3LYP and PBE0), GW approach, and $\Delta$scf method.
At the end, the vibrational modes of the lowest lying isomers are calculated
by using the force constants method and the IR active modes of the systems are identified.
The vibrational spectra are used to calculate the Helmholtz free energy of 
the systems and then to investigate abundance of the nanoclusters at finite temperatures.

\end{abstract}
\maketitle


\section{Introduction}

Transition Metal Dichalcogenides (TMDCs) with chemical formula MX$_2$
are semiconducting compounds made of hexagonal sheets, 
where inside a sheet one layer of metal atom M (Mo, W, Ga, V, Sn, Te) 
is sandwiched between two layers of chalcogen atom X (S, Se, Te).
The weak Van der Waals interaction between MX$_2$ sheets provides the opportunity 
to extract two dimensional (2D) semiconductors with fascinating applications in
supercapacitors \cite{muller2015high,huang2013},
solar cells \cite{ellmer2008,patel2011,tsai2014m,gawale2010,tributsch1978},
lithium batteries \cite{david2014,rouxel1986low,bhandavat2012,xiao2010},
hydrogen evolution in fuel cells \cite{kong2013first,chen2013},
optoelectronics devices
\cite{mak2010, ross2014, lopez2013, wang2012,tian2014novel}
and gas sensor
\cite{wang2012,late2014single}.
Energy gap of these 2D semiconductors lies between 1-2 eV and
compared with graphene, exhibit about 25\% more resistance
to pressure and strain \cite{he2013}.

In addition to the 2D MX$_2$ sheets, other forms of MX$_2$ nanostructures
have also attracted considerable attention \cite{parsapour1996electron,chikan2000relaxation}.
Nanoparticles of these materials exhibit great potential applications
as catalysts in hydrogen evolution reaction and hydrogen desulfurization
\cite{lauritsen2003c,hinnemann2005,mcbride2009,lauritsen2004a,liu20163d,tang2014}.
For instance, MoS$_2$ nanoparticles have been used as catalyst in desulfurization 
processes of raw oil \cite{zhu2013,chianelli2006}.
MoSe$_2$ nanofilms and nanosheets may also catalyze hydrogen evolution reaction 
and regeneration of I$^{-}$ species \cite{lee2014}.
Very recently biosensor applications of MoS$_2$ nanocomposites 
were also reported \cite{zhang2015}.
Moreover, the high chemical stability of these nanostructures
make them very proper lubricants in strongly oxidizing environments
\cite{cizaire2002,gemming2006}.

In this work we investigate structural and electronic properties of 
MoS$_2$ and MoSe$_2$ nanoclusters.
Experimental observations show that these nanoclusters at large sizes 
prefer triangular flat plates shape
\cite{lauritsen2007s,li2007,helveg2000atomic,bertram2006}.
However, the available experimental techniques may not be able 
to identify the atomic structure of very small clusters,
hence accurate computational simulations are valuable
complementary techniques for investigation of the atomic structure 
of very small atomic nanoclusters
\cite{seifert2000,cizaire2002,murugan2007u,murugan2005a}.
Recent proposed algorithm for theoretical structure search are found to be
very powerful and reliable for identifying the most stable and metastable
atomic configuration of crystals and nanostructures
\cite{chuang2006,oganov2008e}.
We will use evolutionary algorithm and first-principles calculations 
to identify the lowest energy atomic configurations
of (MoX$_2$)$_n$ (X = S, Se) nanoclusters for $n<10$.
After finding the lowest energy configurations,
the magic numbers, electronic properties, and vibrational spectra 
of these nanoclusters will be discussed in detail.

\section{Computational details}

Our structure search was performed in the framework of evolutionary algorithm 
developed by Oganov \textit{et al.} and 
featuring local optimization, real-space representation and flexible 
physically motivated variation operators \cite{oganov2006c,lyakhov2013,oganov2011h}. 
This approach starts with a generation of trial structures,
which are randomly generated or taken from some seed structures,
and utilizes an auxiliary total energy code to minimize the energy of the structures.
Then a series of refining and production operators such as 
heredity and mutation are applied to the lowest energy structures of 
the previous generation to create an improved generation of trial structures.
This process is repeated until achieving a stop criteria, 
which is persistence of a specific number of lowest lying (in energy) structures 
in a certain number of generations.
In order to avoid algorithm to be trapped in the local valleys of 
the Born-Oppenheimer energy landscape \cite{glass2006,lyakhov2013},
proper operators for detecting and deleting equivalent structures before 
structural optimization was applied.
For bigger clusters, more diverse structures are possible, and more cases should be investigated.
So, in order to ensure the reliability of the structure search, we increased the value of population size
from 10 to 45, and used stricter stop criteria,
meaning that the number of generations and lowest energy isomers in 
the stop criterion were increased from 15 to 55 and from 8 to 15, respectively.
It should be noted that for generating a new population of trial structures,
during the structure searches, 50\% of the new structures were generated by heredity operator,
10\% by random operator, 10\% by permutation, and 20\% by softmutation operator.
Moreover, the structures obtained from Ref.\cite{murugan2005a} have been added to our structure search,
as seed structures.

The total energy calculations and minimizations of structures were performed in the framework 
of Kohn-Sham density functional theory (DFT) by using the all-electron full-potential method 
implemented in FHI-aims package\cite{blum2009ab,ren2012resolution}. 
This code employs numeric atom-centered orbital (NAO) basis functions 
which are very efficient for computation of non-periodic systems.
In order to increase the computational performance, the geometrical relaxation of 
the clusters was done in four steps: 
initial relaxation with a light basis set, ignoring relativistic and spin-polarization effects,
secondary relaxation by turning on the relativistic effects at scalar level and a tighter basis set, 
third step with addition of spin-polarization effect,
and final relaxation with a higher number of basis functions (tight+tier2).
The relaxation process of all structures was performed down to 
residual atomic forces of less than $10^{-3}$eV/\AA,
while for calculating vibrational frequencies and IR spectra of 
the lowest energy isomers, a maximum atomic force of $10^{-4}$eV/\AA, was considered.

The structural relaxations were performed within Becke-Lee-Yang-Parr (BLYP)\cite{becke1988multicenter, lee1988development} functional,
which, compared with Perdew-Burke-Ernzerhof (PBE) functional, 
sounds to be more accurate for studying molecular systems.
\cite{martin2004,koch2015chemist}.
It is argued that computational error of exchange energy 
within BLYP and PBE functionals are roughly equal,
while correlation energy of molecular systems 
within BLYP is about 4 kcal/mol more accurate
\cite{kurth1999molecular}. 
For more accurate study of the electronic properties, 
the HOMO-LUMO gap of the lowest energy isomers was calculated within four functionals; 
two generalized gradient approximations (GGA), BLYP and PBE,
and two hybrid functionals, namely B3LYP \cite{becne1993densityddotnufunctional} and PBE0 \cite{ernzerhof1999assessment, adamo1999toward}.
Moreover, the many body based G$_0$W$_0$ and the total energy based $\Delta$SCF 
techniques were applied to increase the accuracy of the calculated energy gaps.

\section{Results and discussion}
\label{sec:results}
\subsection{Stable and Metastable Isomers}  

\begin{figure*}
\includegraphics[scale=1]{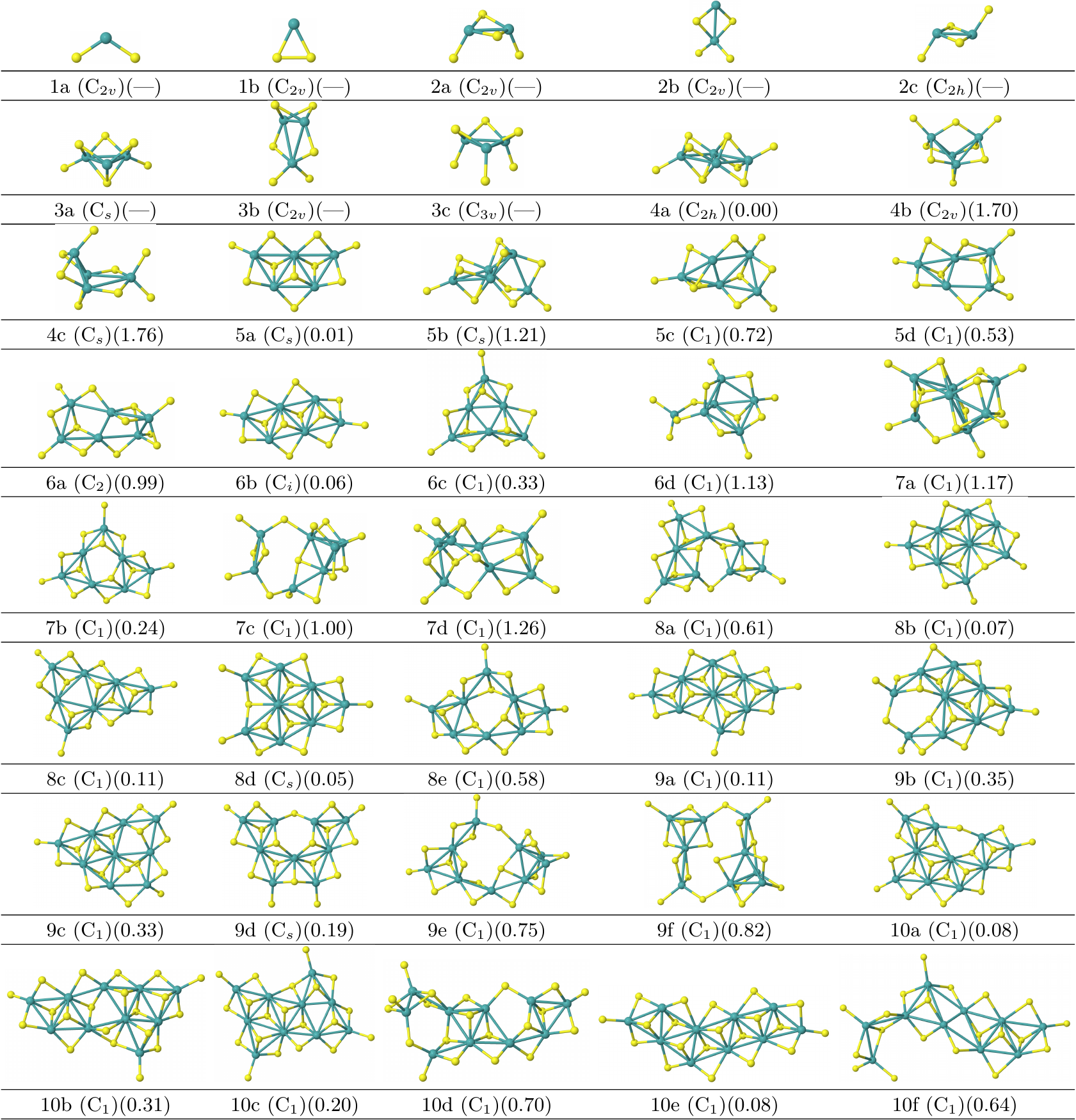}
\caption{\label{struct-MoS2}
 Stable and metastable structures of (MoS$_2$)$_n$ ($n=1-10$) nanoclusters.
 The small yellow (large blue) balls stand for S (Mo) atoms.
 The point groups, written in the parentheses, 
 are determined by using MacMolPlt software \cite{bode1998macmolplt}.
 Obtained Root Mean Square Deviation (RMSD) of molybdenum atoms
 from a perfect flat plane is written in the second parenthesis.
}
\end{figure*}

\begin{figure*}
\includegraphics[scale=1]{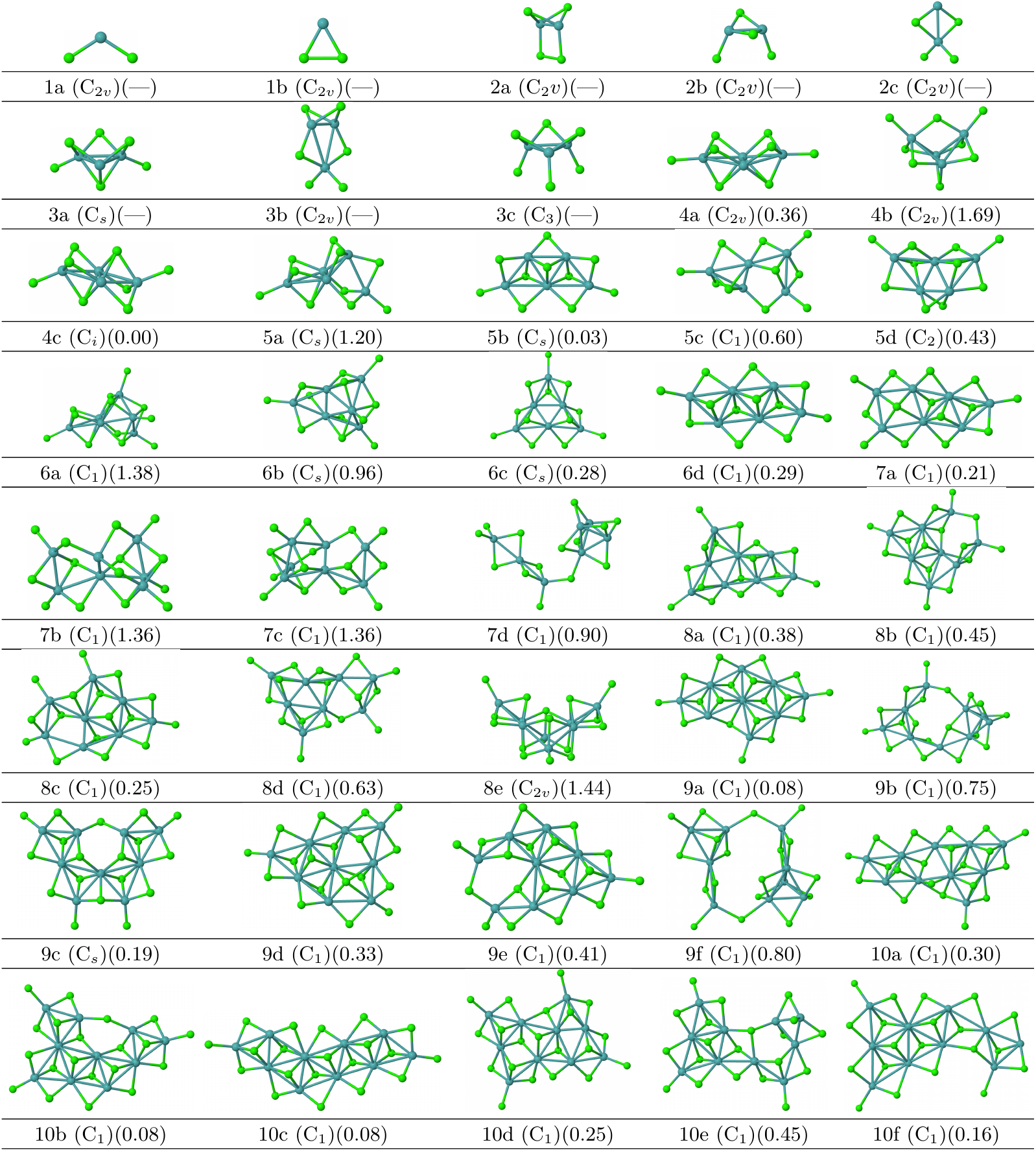}
\caption{\label{struct-MoSe2}
 Stable and metastable structures of (MoSe$_2$)$_n$ ($n=1-10$) nanoclusters.
 The small green (large blue) balls stand for Se (Mo) atoms.
 The point groups, written in the first parentheses, 
 are determined by using MacMolPlt software \cite{bode1998macmolplt}.
 Obtained Root Mean Square Deviation (RMSD) of molybdenum atoms
 from a perfect flat plane is written in the second parenthesis.
}
\end{figure*}

The obtained lowest energy structural isomers of (MoS$_2$)$_n$ and 
(MoSe$_2$)$_n$ ($n=1-10$) clusters are shown in figures \ref{struct-MoS2} 
and \ref{struct-MoSe2}, respectively, sorted by their minimized energy 
within the BLYP functional.

In the size of $n=1$, the equilibrium X-Mo-X angle in MoS$_2$ and MoSe$_2$ 
is 113$^{\rm o}$\ and the equilibrium Mo-X bond length is 2.13\AA\ and 2.27\AA, respectively.
The larger length of Mo-Se bond is due to larger atomic radius of Se, compared with S.
The Mo-S and Mo-Se bond lengths in the ideal MoX$_2$ sheet are 2.45\AA\ and 2.58\AA\, respectively, 
which are approximately 12\% longer than cluster values.
Lower coordination number of atoms in the nanoclusters gives rise to compressed atomic
bonds, compared with the corresponding periodic system.
In contrast to \textcite{murugan2005a},
we found that linear configuration of these clusters are dynamically unstable.

In the second size ($n=2$), a direct Se-Se bond is seen in Mo$_{2}$Se$_{4}$,
while the situation in Mo$_{2}$S$_{4}$ is different.
It may be attributed to the lower electronegativity of Se (2.43, Allen scale),
compared with S (2.58), which gives rise to less ionic charge transfer from Se to Mo
and thus more chemical activity of Se ion, compared with S ion.
It should be noted that the second isomer of \textcite{murugan2005a} for Mo$_{2}$X$_{4}$
coincides with our sixth isomer, evidencing reliability of our structure search.

From the third size ($n=3$), some sandwiched configurations start 
to appear in the lowest energy isomers. 
In these configurations, similar to an ideal MoX$_2$ sheet, a plane of Mo atoms
is sandwiched between layers of X atoms.
The remarkable point is that with increasing the cluster size, 
their tendency to form two dimensional triangular structures increases.
Some examples are 6b, 7b, 8b, and 8c isomers of MoS$_2$, 6c, 6d, 7a, 8a, 8b, and 8c isomers 
of MoSe$_2$, and stable and several metastable isomers of 
sizes 9 and 10.

For better understanding of the bonding properties of the systems,
we follow the conventions of \textcite{murugan2005a} 
to classify the X atoms of MoX$_2$ clusters in three different groups: 
the X atom can bond to 
I. one Mo atom (terminal atom, X$_T$), 
II. two Mo atoms (bridging atom, X$_B$), and
III. three Mo atoms (face capping atom, X$_C$).
Consequently, four kinds of atomic bonds may occur in the systems,
Mo-Mo, Mo-X$_T$, Mo-X$_B$, and Mo-X$_C$ bonds.
The Mo-X$_C$ bond happens at the central part of the clusters,
while Mo-X$_B$ and Mo-X$_T$ bonds appear at the cluster edges and corners.
Some bonding properties of MoS$_2$ and MoSe$_2$ clusters are presented in table \ref{bond}.
Generally, increment in the cluster size, increases the number of Mo-X$_{C}$ bonds.
The average Mo-X$_C$ bond length in MoS$_2$ and MoSe$_2$ clusters is 2.37\AA\
and 2.5\AA, respectively, which is close to the corresponding values
in the ideal sheets.
It is seen that the shortest bond length in all clusters belongs to Mo-X$_T$ bonds.
Lower coordination number of X$_T$ atoms enhances the strength of Mo-X$_T$ bonding,
compared with Mo-X$_B$ and Mo-X$_C$ bondings.

\begin{table}[!htb]
\caption{\label{bond}
  Calculated average bond length $d$ (\AA) and coordination number $n_c$ 
  in the studied clusters.
  In each row, the first (second) line belongs to the MoS$_2$ (MoSe$_2$) clusters.
  The given numbers in the parentheses show the number of corresponding bond in the system.
  The capital letter M stands for Mo atom.}
\begin{ruledtabular}
\begin{tabular}{cccccc}
 size 
 & $d_{M-M}$ & $d_{M-X_T}$ & $d_{M-X_B}$ & $d_{M-X_C}$ & $n_c$ \\ \hline

\multirow{2}{*}{1} 
 &    ---    &   2.13 (2)  &     ---     &    ---      &   1   \\
 &    ---    &   2.27 (2)  &     ---     &    ---      &   1   \\ \hline
\multirow{2}{*}{2}
 &    ---    &   2.14 (2)  &   2.31 (2)  &    ---      &  1.5  \\  
 &    ---    &   2.52 (2)  &   2.52 (2)  &    ---      &  1.5  \\ \hline
\multirow{2}{*}{3}
 &    2.76   &   2.16 (2)  &   2.33 (3)  &   2.34 (1)  &  1.8  \\
 &    2.78   &   2.30 (2)  &   2.46 (3)  &   2.50 (1)  &  1.8  \\ \hline
\multirow{2}{*}{4}
 &    2.85   &   2.15 (2)  &   2.36 (4)  &   2.35 (2)  &  2.0  \\
 &    2.88   &   2.28 (2)  &   2.50 (4)  &   2.67 (2)  &  2.0  \\ \hline
\multirow{2}{*}{5} 
 &    2.90   &   2.15 (2)  &   2.36 (5)  &   2.38 (3)  &  2.1  \\
 &    2.83   &   2.28 (2)  &   2.49 (7)  &   2.64 (1)  &  1.9  \\ \hline
\multirow{2}{*}{6} 
 &    2.96   &   2.14 (4)  &   2.32 (6)  &   2.37 (2)  &  1.8  \\
 &    2.96   &   2.28 (4)  &   2.49 (4)  &   2.55 (4)  &  2.9  \\ \hline
\multirow{2}{*}{$7$}
 &    2.96   &   2.13 (5)  &   2.34 (5)  &   2.44 (4)  &  1.9  \\
 &    2.93   &   2.28 (3)  &   2.53 (7)  &   2.56 (4)  &  2.1  \\ \hline
\multirow{2}{*}{$8$} 
 &   2.89    &   2.15 (3)  &   2.35 (8)  &  2.34 (5)   &  2.1  \\
 &   2.93    &   2.28 (3)  &   2.53 (8)  &  2.56 (5)   &  2.1  \\ \hline
\multirow{2}{*}{$9$} 
 &   2.90    &   2.15 (3)  &   2.37 (7)  &  2.42 (8)   &  2.2  \\
 &   3.05    &   2.28 (3)  &   2.51 (7)  &  2.55 (8)   &  2.3  \\ \hline
\multirow{2}{*}{$10$} 
 &   3.03    &   2.15 (3)  &   2.38 (9)  &  2.31 (8)   &  2.6  \\
 &   2.99    &   2.28 (3)  &   2.51 (9)  &  2.55 (8)   &  2.2  \\
\end{tabular}
\end{ruledtabular}
\end{table}

For more accurate identification of the planar isomers, 
the Root Mean Square Deviation (RMSD) of Mo atoms from a perfect planar geometry
for some lowest energy isomers of the clusters is calculated 
and listed in Figs.~\ref{struct-MoS2} and \ref{struct-MoSe2}.
It is generally concluded that, unlike the most stable isomers 
of MoS$_2$ in previous studies\cite{murugan2005a,murugan2007u} which 
involve polyhedral core of Mo atoms covered with sulfur atoms,
sandwiched planar configurations are favored by MoS$_2$ and MoSe$_2$ nanoclusters. 
Experimental observations in larger MoX$_2$ nanoclusters ($n=10-100$) 
confirm favorability of flat triangular shapes in these systems \cite{bertram2006}.
However, it should be noted that our predicted flat configurations are more
similar to the 1T structure of TMDC sheets, which is metastable compared with 
the stable 2H structure of these 2D compounds.
Observation of 1T configuration is attributed to the fixed stoichiometry of
the investigated nanoclusters,
while stabilizing 2H triangular clusters requires extra chalcogen atoms 
at the edges \cite{lauritsen2007size}. 
While direct structural observation of very small clusters is hardly feasible,
our accurate structure search provides reliable evidence for favorability of 
flat triangular configurations even in very small MoS$_2$ and MoSe$_2$ nanoclusters.
In contrast to carbon clusters which prefer cage-like or fullerene-like 
structures to saturate their surface dangling bonds,
MoX$_2$ nanoclusters tend to form planar structures 
with possible extra X atoms at the edges and corners 
to saturate dangling bonds \cite{bertram2006}.

Comparing the stable and metastable isomers of MoS$_2$ and MoSe$_2$
with those reported for WS$_2$ nanoclusters \cite{hafizi2016},
clarifies that from structural point of view,
MoS$_2$ isomers are more similar to MoSe$_2$ isomers, compared with WS$_2$.
This results reflects the importance role of transition metals in determining 
the stable geometry of TMDCs nanoclusters.
We also noticed that, WS$_2$ nanoclusters exhibit less tendency to planner configurations, 
because even at larger sizes, WS$_2$ nanoclusters are not quite planner \cite{hafizi2016}.

\begin{table}
\caption{\label{deltaE}
 Obtained energy orders for more important isomers of MoS$_2$ and MoSe$_2$ 
 nanoclusters within PBE and BLYP.
 The energy difference (eV) of metastable isomers with the lowest energy isomer
 is given in the parenthesis.}
\begin{ruledtabular}
\begin{tabular}{ccccc}
 \multicolumn{2}{c}{MoS$_2$} &&  \multicolumn{2}{c}{MoSe$_2$} \\
\cline{1-2}\cline{4-5}                   
      BLYP    &      PBE     &&      BLYP     &      PBE      \\\hline
                               
  2a~~~~~~~~~ &  2a~~~~~~~~~ &&  2a~~~~~~~~~  &  2a~~~~~~~~~  \\  
  2b~(0.31)   &  2b~(0.32)   &&  2b~(0.32)    &  2b~(0.55)    \\  
  2c~(0.33)   &  2c~(0.34)   &&  2c~(0.52)    &  2c~(0.71)    \\\hline
  3a~~~~~~~~~ &  3a~~~~~~~~~ &&  3a~~~~~~~~~  &  3a~~~~~~~~~  \\  
  3b~(0.48)   &  3b~(0.79)   &&  3b~(0.24)    &  3b~(0.55)    \\  
  3c~(0.66)   &  3c~(1.05)   &&  3c~(0.47)    &  3c~(0.89)    \\\hline
  4a~~~~~~~~~ &  4a~~~~~~~~~ &&  4a~~~~~~~~~  &  4c~~~~~~~~~  \\  
  4b~(0.07)   &  4b~(0.46)   &&  4b~(0.01)    &  4b~(0.19)    \\  
  4c~(0.15)   &  4c~(0.80)   &&  4c~(0.13)    &  4a~(0.19)    \\\hline
  5a~~~~~~~~~ &  5a~~~~~~~~~ &&  5a~~~~~~~~~  &  5b~~~~~~~~~  \\  
  5b~(0.22)   &  5b~(0.64)   &&  5b~(0.07)    &  5a~(0.32)    \\  
  5c~(0.26)   &  5c~(0.79)   &&  5c~(0.28)    &  5d~(0.57)    \\  
  5d~(0.37)   &  5d~(1.01)   &&  5d~(0.30)    &  5c~(0.81)    \\\hline
  6a~~~~~~~~~ &  6b~~~~~~~~~ &&  6a~~~~~~~~~  &  6a~~~~~~~~~  \\  
  6b~(0.25)   &  6c~(0.73)   &&  6b~(0.07)    &  6b~(0.04)    \\  
  6c~(0.40)   &  6a~(0.79)   &&  6c~(0.09)    &  6d~(0.15)    \\  
  6d~(0.45)   &  6e~(0.90)   &&  6d~(0.42)    &  6c~(0.10)    \\ 
  6e~(0.47)   &  6d~(1.49)   &&               &               \\\hline
  7a~~~~~~~~~ &  7b~~~~~~~~~ &&  7a~~~~~~~~~  &  7a~~~~~~~~~  \\  
  7b~(0.13)   &  7a~(0.39)   &&  7b~(0.17)    &  7e~(0.71)    \\  
  7c~(0.25)   &  7d~(1.17)   &&  7c~(0.18)    &  7c~(0.72)    \\  
  7d~(0.26)   &  7c~(1.23)   &&  7d~(0.20)    &  7b~(0.72)    \\  
  7e~(0.50)   &  7e~(1.61)   &&  7e~(0.21)    &  7d~(1.33)    \\\hline
  8a~~~~~~~~~ &  8b~~~~~~~~~ &&  8a~~~~~~~~~  &  8a~~~~~~~~~  \\  
  8b~(0.01)   &  8d~(0.29)   &&  8b~(0.14)    &  8c~(0.39)    \\  
  8c~(0.25)   &  8a~(0.54)   &&  8c~(0.22)    &  8b~(0.49)    \\  
  8d~(0.30)   &  8c~(0.64)   &&  8d~(0.26)    &  8d~(0.82)    \\  
  8e~(0.37)   &  8e~(1.270   &&  8e~(0.28)    &  8e~(1.05)    \\\hline
  9a~~~~~~~~~ &  9a~~~~~~~~~ &&  9a~~~~~~~~~  &  9a~~~~~~~~~  \\  
  9b~(0.34)   &  9c~(0.55)   &&  9b~(0.38)    &  9d~(0.51)    \\  
  9c~(0.39)   &  9b~(0.74)   &&  9c~(0.39)    &  9e~(0.85)    \\  
  9d~(0.49)   &  9d~(1.01)   &&  9d~(0.41)    &  9c~(0.97)    \\  
  9e~(0.59)   &  9e~(1.76)   &&  9e~(0.41)    &  9b~(1.60)    \\\hline
  10a~~~~~~~~ & 10a~~~~~~~~~ &&  10a~~~~~~~~~ &  10a~~~~~~~~~ \\  
  10b~(0.22)  &  10b~(0.19)  &&  10b~(0.09)   &  10b~(0.12)   \\  
  10c~(0.63)  &  10e~(0.66)  &&  10c~(0.28)   &  10c~(0.25)   \\  
  10d~(0.70)  &  10c~(0.94)  &&  10d~(0.33)   &  10d~(0.60)   \\  
  10e~(0.72)  &  10d~(1.74)  &&  10e~(0.78)   &  10e~(1.48)   \\ 
\end{tabular}
\end{ruledtabular}
\end{table}

In order to check the effect of exchange-correlation functional
on atomic configuration of MoX$_2$ clusters,
some lowest energy isomers of all nanoclusters were relaxed 
within the PBE and B3LYP functionals.
The obtained new energy orders within these two functionals
are compared with BLYP in table~\ref{deltaE}.
In the case of MoS$_2$, we observe that PBE displaces
the first isomer of 6, 7 and 8 clusters, 
while in other sizes, most stable isomers are the same within PBE and BLYP.
A more precise consideration show that the lowest energy isomers of 
6, 7, and 8 clusters within BLYP are not quite flat and triangular,
it seems that PBE is trying to stabilize more triangular structures.
From another point of view, more flat isomers with more number of 
Mo-S$_C$ bonds are better favored within PBE.
For example in the size of $n=6$, PBE favors 6b isomer 
which has very small RMSD value (Figs.~\ref{struct-MoS2})
and more number of Mo-S$_C$ bonds, compared with other isomers.
Then the 6c, 6a, and 6d isomers with respectively 3, 2, and 1 Mo-S$_C$ bond, 
occupy the next places in the obtained energy order within PBE.

The same trend is observed in MoSe$_2$ clusters (table~\ref{deltaE}). 
Compared with BLYP, PBE only displaces stable isomer of clusters 4 and 5.
Some metastable structures are also displaced according to 
the above-mentioned trend. 
For example, in the size of $n=9$, the first isomer is not changed within PBE,
compared with BLYP, but the isomers 9d, 9e, and 9c with seven Mo-Se$_{C}$ bonds 
are more stable than 9b, within PBE.
The isomer 9b has the lowest number of Mo-Se$_C$ bonds in this group.
On the other hand, among the three isomers with the same number of Mo-Se$_{C}$ bonds,
those with a more triangular shape are more stable.

Hence the general statement is that PBE, compared with BLYP, 
favors more flat triangular structures of MoX$_2$ nanoclusters.
It might be related to the better performance of PBE functional 
in description of periodic structures \cite{martin2004}, 
because flat structures are closest cluster
configurations to the arrangement of atoms in the ideal
octahedral coordination of 1T-MoX$_2$ sheets.
We also re-calculated several isomers within the B3LYP hybrid functional 
and found that stability order of isomers with this functional
is very similar to BLYP.
The same feature has been observed for WS$_2$ nanoclusters \cite{hafizi2016}.

It is already observed that nanoclusters of a non-magnetic material 
may exhibit magnetism \cite{hafizi2016,murugan2005a}.
In order to find the stable magnetic state of the clusters,
their total energy were minimized by considering different initial magnetization
and then the lowest energy state was reported as the stable magnetic state.
It was found that the lowest energy isomer of the first size ($n=1$) 
of MoS$_2$ and MoSe$_2$ clusters have a total spin moment of 2\,$\mu_B$, 
within both BLYP and PBE.
The lowest energy isomer of MoSe$_2$ clusters at all other investigated sizes
are nonmagnetic, while those of MoS$_2$ clusters at the second and sixth sizes
exhibit a total spin moment of 2\,$\mu_B$, within both BLYP and PBE.

\begin{figure*}
\includegraphics[scale=0.80]{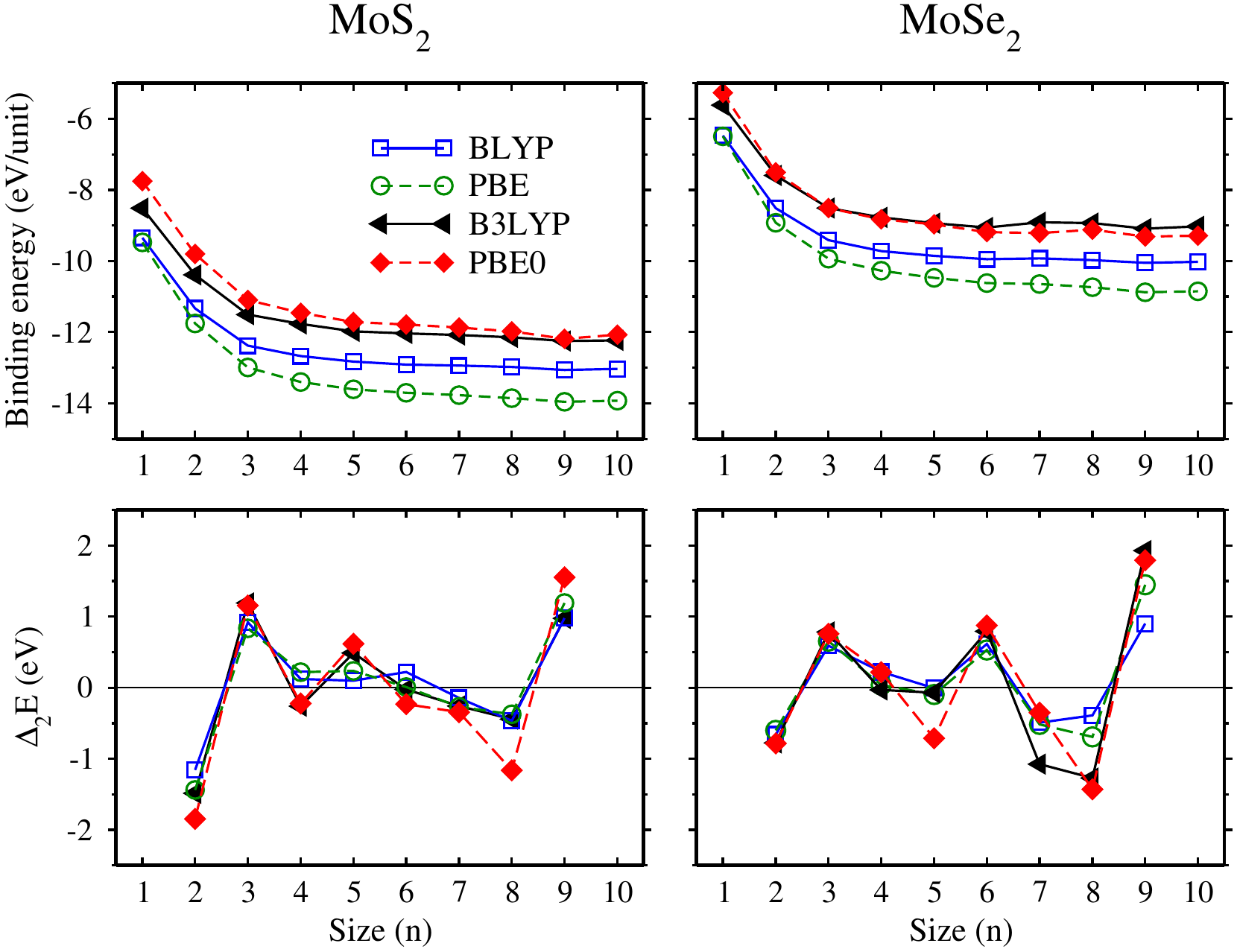}
\caption{\label{energy}
 Calculated binding energy (eV/unit) and second order difference in energy,
 $\Delta_2E$ (eV) of the lowest energy isomers of MoS$_2$ and  MoSe$_2$ 
 nanoclusters within the BLYP, PBE, PBE0, and B3LYP functionals.}
\end{figure*}

\subsection{Relative Stability of Nanoclusters}  

In order to address the stability of the nanoclusters, 
their binding energy (BE) per MoX$_2$ unit was calculated, as follows:
\begin{equation}
BE(n)=\frac{E_{tot}(n)-n(E_{at}({\rm Mo})+2E_{at}(S))}{n}
\label{equ-BE}
\end{equation}
where $E_{tot}(n)$ is the minimized total energy of the most stable isomer of 
the cluster of size $n$ and $E_{at}$(X) is the total energy of a free X atom.
The binding energies were calculated within four different 
functionals and presented in Fig.~\ref{energy}.
According to this diagram, the binding energy (BE) is a monotonic increment with size, 
that indicating more stability of larger clusters, 
which is due to the reduction of the relative number of surface dangling bonds.
Taking into account the tendency of the lowest lying isomers to have 
planar/semi-planar configurations,
it is expected that BE plots converge to the binding energy of 
ideal MoS$_2$ and MoSe$_2$ sheets, 
which were calculated 13.7\,eV and 10.63\,eV within BLYP and 15.11\,eV and 
11.91\,eV within PBE, respectively.
There is a local minimum at $n=9$ in all functionals,
which indicates more relative stability of this size compared with neighboring sizes. 
In the case of MoSe$_2$, another local minimum is visible at $n = 6$,
while $n = 7,8$ occur on a local bulge of the BE plot, 
indicating lower relative stability of the 7th and 8th clusters.

For more accurate description of the relative stability of the systems
and identifying the small magic sizes of the nanoclusters, 
the second-order difference in energy is defined as follows:

\begin{equation}
\Delta_{2}E=E_{tot}(n+1)+E_{tot}(n-1)-2E_{tot}(n)
\label{equ-delta2}
\end{equation}
This parameter is conventionally expected to be comparable with 
the mass spectrometry measurements on nanoclusters. 
The reason is that from a physical point of view, a positive value of $\Delta_{2}E$
indicates higher relative stability and consequently more abundance of 
the corresponding cluster, compared with the neighboring sizes. 
Hence the local peaks of the $\Delta_{2}E$ plot is expected 
to happen at the magic sizes of the clusters.
The calculated $\Delta_{2}E$ values for MoX$_{2}$ clusters within 
four functionals are shown in Fig.~\ref{energy}.
It is seen that the $n=3,9$ sizes of MoS$_{2}$ nanoclusters exhibit
high abundance within all functionals,
while the 5th cluster shows good relative stability only within hybrid functionals.
On the other hand, the sizes of $n=2,8$ show very low abundance within all functionals.
These findings contradict with previous reports on 
the magic sizes ($n=2,4,6$) of small MoS$_{2}$ clusters \cite{murugan2005a}. 
This is due to the different stable isomers found in this work.
The lowest energy isomers in the work of \textcite{murugan2005a}
usually coincide with our high energy metastable isomers,
found in our comprehensive systematic structure search.
In the case of MoSe$_2$ nanoclusters, all functionals predict 
magic sizes of $n=3,6,9$,
while the sizes of $n=2,7,8$ exhibit very low relative stability.
The large value of $\Delta_{2}E$ at the 9th size of both systems
along with the observed local minimum at this size in the binding energy plots,
confirm that $n=9$ is an important magic size of MoS$_2$ and MoSe$_2$ nanoclusters.
An interesting point is that, except for the lowest magic size ($n=3$),
other magic sizes occur on the systems with higher relative
number of Mo-Se$_C$ bonds, compared to neighboring sizes (see table \ref{bond}).

\begin{figure}
\includegraphics[scale=0.8]{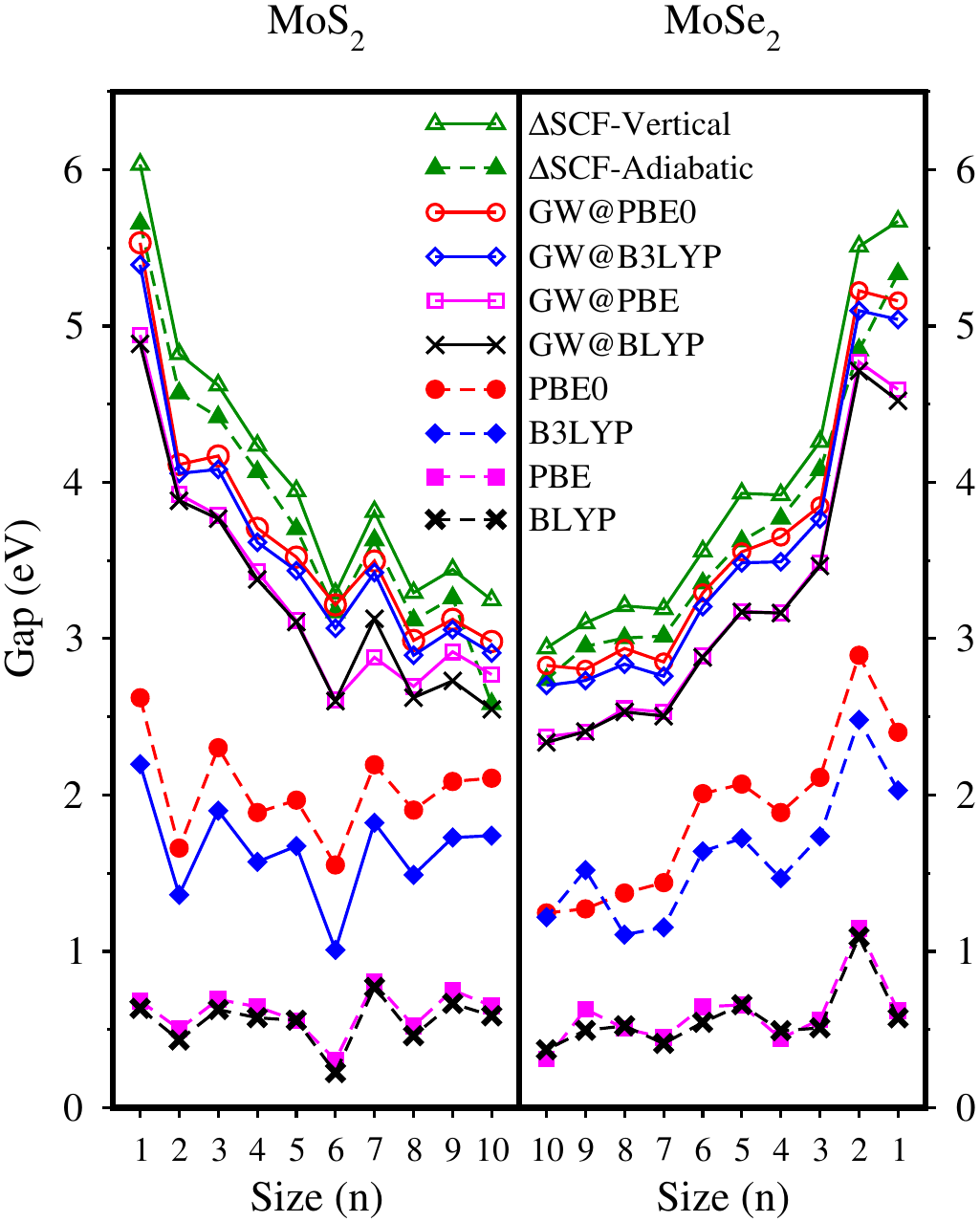}
\caption{\label{gap}
 Obtained energy gap of MoX$_2$ nanoclusters 
 as a function of size, by using ten different methods.
}
\end{figure}

\subsection{Energy Gap}  

The HOMO-LUMO energy gap of the lowest energy isomers of MoX$_2$ nanoclusters
was calculated within the BLYP, PBE, B3LYP, and PBE0 functionals (Fig.~\ref{gap}).
It is seen that PBE and BLYP give very similar and rather uniform energy gaps,
while hybrid B3LYP and PBE0 functionals display significantly enhanced values.
Despite the success of LDA/GGA functionals in predicting structural properties, 
presence of self-interaction error (SIE) gives rise to significantly underestimated 
energy gaps within these functionals \cite{koch2015chemist,ke2011all}.
On the other hand, hybrids functionals partially resolve the SIE problem
through combination of Hartree-Fock exchange with the GGA one,
thus enhancing the energy gap compared with GGA \cite{koch2015chemist}.
One of the most effective methods for correcting the DFT HOMO-LUMO gap
is the many-body based GW approximation 
\cite{caruso2012,marom2012,friedrich2006,rinke2005,körbel2014}.
In this method, a week screened coulomb interaction is switched on between 
the fictitious Kohn-Sham particles to perturbatively obtain the quasiparticle 
spectra of the system by using the Green's function technique.
We performed non-self-consistent G$_0$W$_0$ calculation, 
where the mean field Green's function, G$_0$, and the screened Coulomb interaction, W$_0$,
are determined from the Kohn-Sham eigenvalues and eigenvectors.
Therefore, the accuracy of the non-self-consistent G$_0$W$_0$ calculation 
are sensitive to the starting point exchange-correlation functional.
By proper selection of the starting point,
the accuracy of the G$_0$W$_0$ results approaches that of 
the full-self-consistent GW \cite{caruso2012}. 
It is argued that hybrid functionals are better starting points,
compared with LDA, GGA, and Hartree-Fock method for G$_0$W$_0$ calculation.
\cite{körbel2014,bruneval2012,knight2016}
However, it is observed that the calculated energy gaps of 4d and 5d materials by using 
the G0W0@PBE0 method show very small deviation from experiment \cite{körbel2014}.

Figure~\ref{gap} displays the calculated energy gaps within different functionals
after application of the many-body G$_0$W$_0$ correction.
As expected, the G$_0$W$_0$ correction significantly enhances 
the value of the band gaps within all functionals.
The enhancement is more pronounced in the smaller clusters,
which is due to the lower electronic screening in these systems.
As mentioned before, in the G$_0$W$_0$ method, a screened coulomb
interaction is switched on between the Kohn-Sham particles.
Therefore, lower screening in the smaller clusters enhances
the effect of G$_0$W$_0$ correction in the energy gap.
Moreover, it is seen that the G$_0$W$_0$ correction decreases the difference
between the band gap within hybrid and GGA functionals,
in such a way that after G$_0$W$_0$, PBE0 and B3LYP give very similar energy gaps.
Hence, we deduced that B3LYP may also be a proper starting point for
G$_0$W$_0$ calculations in the MoX$_2$ clusters.
It is observed that MoS$_2$ nanoclusters have greater energy gap,
compared with MoSe$_2$ nanoclusters,
which is due to the stronger M-X bonding in 
the MoS$_2$ nanoclusters (table~\ref{bond}).
The same trend is observed in the ideal MoS$_2$ and MoSe$_2$ sheets.
From a qualitative point of view, energy gap is a measure of 
the chemical hardness of the system \cite{kohn1996density}.
The larger the band gap, the more energy required to
disrupt the electronic structure of the system, and the
lower chemical activity of the system.

An alternative approach for calculation of energy gap is $\Delta$SCF method.
In this method, an electron is added to and subtracted from the neutral system, and 
then total energy of the ionized systems are computed
to obtain the energy gap as follows:
\begin{equation}
gap=E_{tot}(cation)+E_{tot}(anion)-2E_{tot}(neutral)
\label{deltascf}
\end{equation}
If the energy of the ionized system is calculated without relaxation,
(exactly at the relaxed geometry of the neutral cluster), 
the method is called vertical $\Delta$SCF,
while in the adiabatic version of the method, the ionized systems are fully relaxed.
This method gives very accurate energy gap for molecular systems \cite{lee2015}.
We used adiabatic and vertical $\Delta$SCF technique along with the BLYP
functional to calculate energy gap of MoX$_2$ nanoclusters (Fig.~\ref{gap}).
It is seen that adiabatic $\Delta$SCF gives energy gaps
very close to the G$_0$W$_0$@PBE0 results, 
while vertical $\Delta$SCF predicts slightly larger energy gaps.

The obtained energy gap of nanoclusters displays a smooth decrease
as a function of the cluster size, which may be explained 
by the quantum confinement effect.
Quantum confinement in nanostructures enhances their energy gap 
with respect to the periodic structures.
Since our flat stable isomers are approaching the 1T phase of 
the corresponding periodic systems, which is found to be metallic,
the energy gap of the flat stable isomers is expected to converge 
to zero at large sizes.

\subsection{Vibrational Frequency}  

The dynamical properties of the lowest lying isomers were investigated 
by identifying the vibrational modes of the systems in 
the framework of the force constant method. 
In this method, a finite displacement ($\delta$) is applied to the atomic positions 
of the fully relaxed structures and then the force set on the atoms
is accurately computed with an error of less than 10$^{-4}$\,eV/\AA. 
The obtained Hessian matrix is then diagonalized to obtain 
the vibrational modes and frequencies of the clusters.
It should be noted that the vibrational properties are 
calculated within the BLYP functional.
The absence of imaginary vibrational modes in the systems  
indicates dynamical stability of the lowest lying isomers. 
All presented meta-stable isomers are also expected to be dynamically stable,
because these configurations are of very low symmetry, obtained through
accurate structural relaxation.
Generally, it is very unlikely that a low symmetry configuration of atoms
traps on a saddle point of the potential energy surface after full atomic relaxation.
The obtained vibrational modes are in the frequency range of $0-390$ and $0-510$\,cm$^{-1}$
for MoS$_2$ and MoSe$_2$ nanoclusters, respectively.
The observed difference is due to the somewhat stronger/shorter 
average bonding in MoS$_2$ with respect to MoSe$_2$.

\begin{figure}
\includegraphics[scale=0.8]{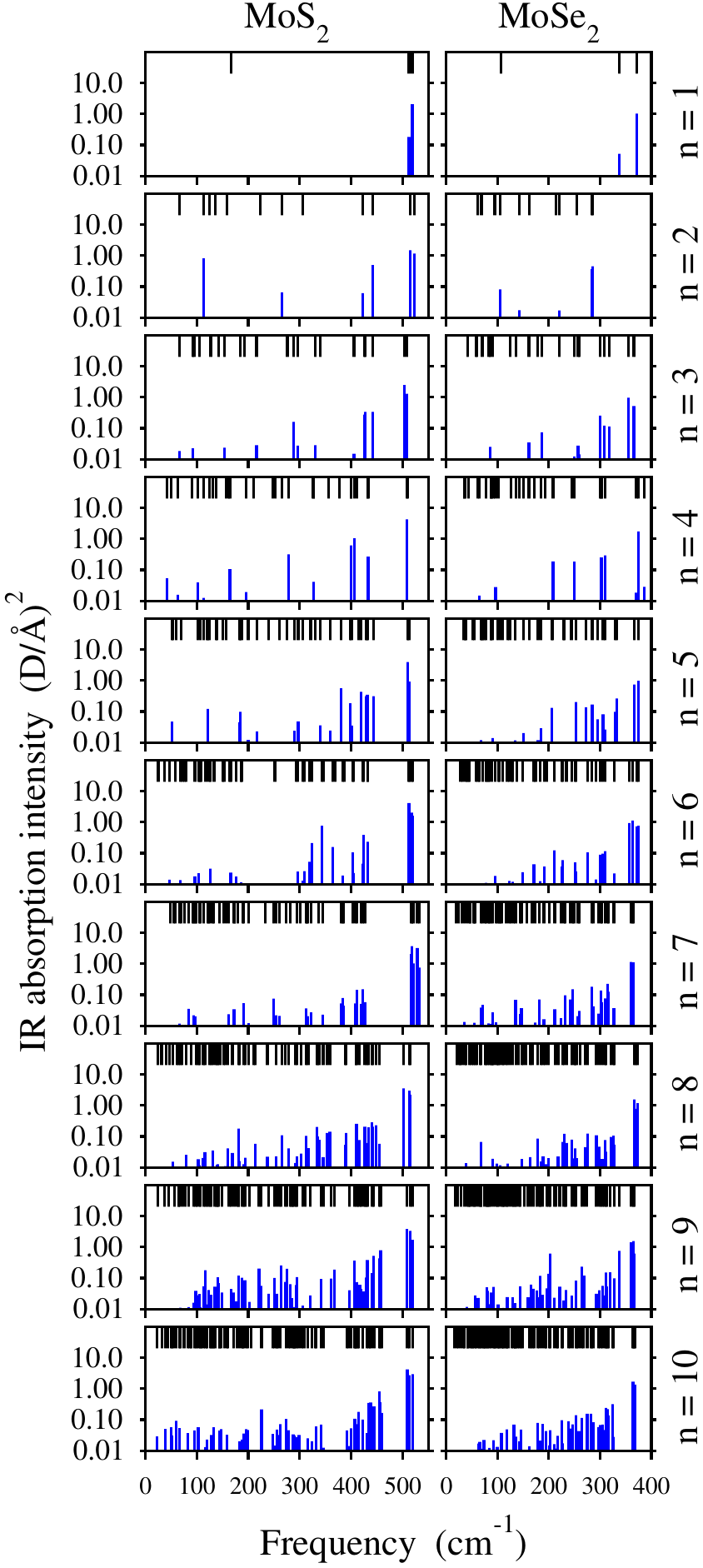}
\caption{\label{ir}
 Computed IR spectra of the lowest energy isomers of 
 (MoS$_2$)$_n$ and (MoSe$_2$)$_n$ nanoclusters.
 The short vertical lines in the upper part of plots
 show the vibrational frequencies of the systems.
}
\end{figure}

IR intensity of the lowest energy isomers at their vibrational modes
are computed by considering the variation of cluster dipole moments  
along the vibrational modes.
The obtained IR intensities, presented in Fig.~\ref{ir},
shows that in practice which IR photons are better absorbed and 
converted to thermal vibration in the systems.
It is seen that the highest IR absorption happens at the high frequency modes,
which corresponds to the strongest Mo-X bonds in the clusters.
It was argued that among various Mo-X bonds, higher degree of hybridization 
happens between Mo and terminal X$_{T}$ atom, 
and hence terminal Mo-X bonds are stronger and shorter (table~\ref{bond}).
As a result, these bonds are the origin of 
the hard vibrational modes of the clusters.
Since these bonds are on the edges of the clusters, 
their dipole moment is more flexible and hence they have
more IR absorption intensity.
The obtained IR spectra indicate that the major part of the IR photons
are thermally adsorbed by the hard edge bonds of the nanoclusters.
(please note the vertical logarithmic scale in Fig.~\ref{ir}).
The other vibrational modes, distributed below the high frequency mode,
are mainly originated form the collective vibration of 
Mo-X$_B$ and Mo-X$_C$ bonds in the systems.
Vibration of these bonds creates less polarization in the systems,
compared with the terminal bonds,
and hence their IR intensity is considerably lower.
With increasing the cluster size, the number of Mo-X$_B$ and Mo-X$_C$ bonds
is increased in the systems and hence more IR active modes
are appeared in the IR spectra.

\begin{figure}
\centering
\includegraphics[scale=0.8]{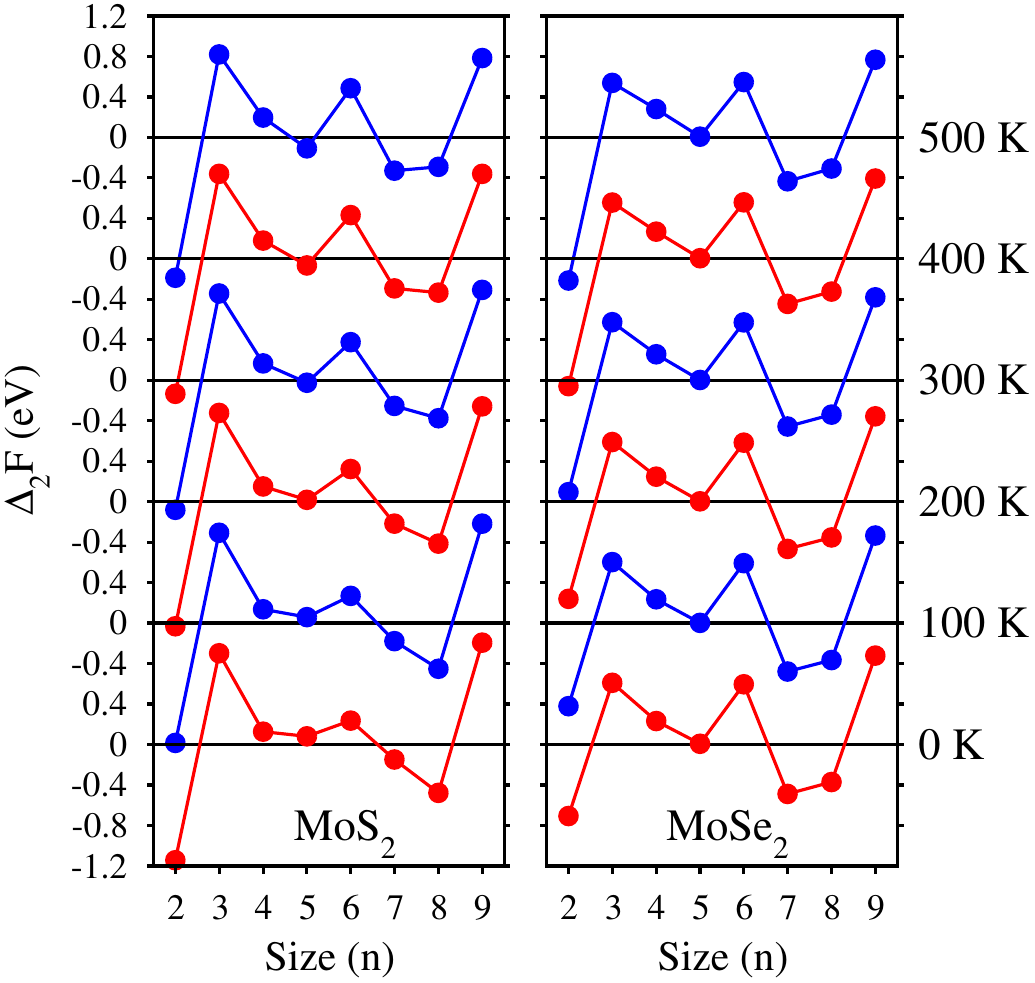}
\caption{\label{d2F}
Second-order difference in vibrational Helmholtz free energy ($\Delta_2$F) 
as a function of the cluster size at the temperature range of $0-500$\,K.}
\end{figure}

In order to investigate the influence of thermal vibrations on 
relative stability of the clusters at elevated temperatures, 
the vibrational Helmholtz free energy ($F_{vib}$) of the clusters
was calculated as follows:
\begin{equation}
F_{vib} = E_{tot} - E_{at} + \frac{1}{2}\sum_{i} \varepsilon_i +
          k_BT\sum_{i} ln(1-e^{\varepsilon_{i}/k_BT})
\label{equ-Fvib}
\end{equation}
where $E_{tot}$ is the minimized total electronic energy of the lowest energy isomers, 
$E_{at}$ is the sum of the free atom energies of the system, 
$i$ runs over the number of vibrational modes, 
$\varepsilon_{i}$ is the energy of the $i$th mode, 
$k_{B}$ is the Boltzmann constant, $T$ is the Kelvin temperature.
The second order difference in Helmholtz free energy ($\Delta_{2}F$) 
was calculated in different temperatures, 
ranging from zero to 500\,K (Fig.~\ref{d2F}). 
In the case of MoS$_2$, we observe that thermal effects enhance $\Delta_{2}F$ 
of even clusters (except $n=2$), which is more pronounced in the 6th cluster. 
It is likely attributed to the more low-energy vibrational modes of the 6th cluster,
compared with the neighboring systems (Fig.~\ref{ir}). 
With increasing temperature, these low-energy vibrational modes give rise 
to faster entropy increase and consequently faster free energy decrease of 
the system in such a way that above room temperature, 
$n=6$ becomes an important magic number of MoS$_2$ nanoclusters.
In the case of MoSe$_2$ nanoclusters, slight thermal effects 
is seen up to 500\,K and only cluster of size $n=4$ shows 
slightly enhanced relative stability at above room temperatures.

In addition to the relative stability of the isomers, 
thermal effects may also influence the lowest energy 
atomic configuration of the clusters.
In other words, increasing temperature may change the stability order of
the lowest energy isomers of a cluster and induce a structural transition
from one isomer to another one, similar to structural transitions reported
in MoX$_2$ single layers \cite{ambrosi2015,duerloo2014}.
In order to address this issue, we focus on the $n=4,6$ sizes of MoS$_2$ and 
$n=4,5,6$ of MoSe$_2$ nanoclusters, because in these systems the first metastable
isomer has a very close energy to the lowest energy isomer (table~\ref{deltaE}) 
and hence thermal effects are expected to be more visible in these cases.
The Helmholtz free energy difference between the stable and the first metastable 
configurations of these systems was calculated at three different temperatures,
presented in table.~\ref{transition}.
The results indicate absence of any structural transition in these
systems up to 500\,K, because the stable isomer remains lower in energy
with respect to the first metastable isomer at temperatures below 500\,K.
Only in the 4th size of MoS$_2$, we observe that the second isomer is 
getting very close to the first isomer at 500\,K and hence one may expect
a structural transition above this temperature.
However, we did not continue our free energy calculations to the higher
temperatures, because our vibrational studies are in the harmonic
approximation limit, which might not be valid at very high temperatures.

\begin{table}
\caption{\label{transition}
 Calculated difference between the total energy $\Delta$E (meV)
 and the Helmholtz free energy $\Delta$F (meV) of the stable 
 and the first metastable isomer of MoS$_{2}$ and MoSe$_{2}$ nanoclusters
 at some specific sizes $n$, at different temperatures.}
\begin{ruledtabular} 
\begin{tabular}{lccccc} 
cluster  & size & $\Delta$E & $\Delta$F & $\Delta$F & $\Delta$F \\ 
         &      &    (0K)   &   (0K)    &   (300K)  &   (500K)  \\ \hline 
MoS$_2$  &   4  &     73    &    49     &     29    &      5    \\ 
         &   8  &     11    &    18     &     73    &    115    \\
MoSe$_2$ &   4  &     7     &     3     &      9    &     45    \\ 
         &   5  &    71     &    79     &     99    &    163    \\ 
         &   6  &    68     &    30     &     50    &    103    \\ 
\end{tabular}
\end{ruledtabular}
\end{table}

\section{Conclusions}

In this paper, we used evolutionary algorithm along with full potential
density functional calculations to identify the stable and metastable 
structures of (MoS$_2$)$_n$ and (MoSe$_2$)$_n$ ($n=1-10$) nanoclusters.
The structure search was done within the BLYP functional and
then the energy order of the lower energy isomers was verified within PBE.
It was argued that the clusters favor sandwiched planar triangular structures 
even in small sizes, similar to the ideal sheets of these systems.
The results show that the lowest energy isomer is usually the same within BLYP and PBE,
while PBE prefers more flat metastable structures with higher number 
of face capping bonds.
The second-order difference in minimized energy of the systems 
indicate that the robust magic sizes of (MoS$_2$)$_n$ and (MoSe$_2$)$_n$ 
nanoclusters are $n=3,9$ and $n=3,6,9$, respectively. 
It was argued that vibrational excitations at finite temperatures enhance 
relative stability of the 6th size of MoS$_2$ nanoclusters to become a magic size 
at above room temperatures,
while the magic sizes of MoS$_2$ nanoclusters remain unchanged up to 500\,K.
The HOMO-LUMO energy gap of the lowest energy isomers 
was obtained by using the hybrid PBE0 and B3LYP functionals, 
the many body based G$_0$W$_0$ technique,
and the total energy based $\Delta$SCF method.
It was argued that G$_0$W$_0$@PBE0 is expected to give 
reliable energy gaps for the investigated nanoclusters.
The energy gap of the systems was found to increase with decreasing
the cluster size, because of the quantum confinement effect.
Finally, we calculated the vibrational spectra and the IR active
modes of the lowest energy isomers.
It was revealed that the most IR absorption happens at a frequency of 
about 510 and 370\,cm$^{-1}$ in MoS$_2$ and MoSe$_2$ nanoclusters, respectively.
These IR photons are mainly absorbed by the strong terminal
Mo-S and Mo-Se bonds located at the corners of the nanoclusters.

\begin{acknowledgments}
This work was jointly supported by the Vice Chancellor
of Isfahan University of Technology (IUT) in Research Affairs and
Centre of Excellence for Applied Nanotechnology.
It should also be acknowledged that the first two authors
have equal contributions in this paper.
\end{acknowledgments}

\bibliography{TMDC}

\end{document}